\documentclass{aa}
\usepackage{graphicx}
\usepackage{longtable}
\usepackage{txfonts}

\newcommand{\grp}    {${\rlap.}^{\circ}$}
\newcommand{\pri}    {${\rlap.}^{\prime \prime}$}

\newcommand{\ltsima} {$\; \buildrel < \over \sim \;$}
\newcommand{\simlt}  {\lower.5ex\hbox{\ltsima}}            
\newcommand{\gtsima} {$\; \buildrel > \over \sim \;$}
\newcommand{\simgt}  {\lower.5ex\hbox{\gtsima}}            

\begin{document}

\title{Radio detections towards unidentified variable EGRET sources}

\author{
J.M. Paredes\inst{1}
\and J.Mart\'{\i}\inst{2,7}
\and C.H. Ishwara-Chandra\inst{3}
\and D.F. Torres\inst{4}
\and G.E. Romero\inst{5,8}\thanks{Member of CONICET}
\and J.A. Combi\inst{2,7}
\and V. Bosch-Ramon\inst{6}
\and A.J. Mu\~noz-Arjonilla\inst{2,7}
\and J.R. S\'anchez-Sutil\inst{7}
}

\offprints{J.M. Paredes}

\institute{
Departament d'Astronomia i Meteorologia and Institut de Ci\`encies del Cosmos (ICC), 
Universitat de Barcelona (UB/IEEC), Mart\'{\i} i Franqu\`es 1, 08028 Barcelona, Spain \\
\email{jmparedes@ub.edu}
\and Departamento de F\'{\i}sica, EPS,  
Universidad de Ja\'en, Campus Las Lagunillas s/n, Edif. A3, 23071 Ja\'en, Spain \\
\email{jmarti@ujaen.es, jcombi@ujaen.es, ajmunoz@ujaen.es} 
\and NCRA, TIFR, Post Bag 3, Ganeshkhind, Pune-411 007, India \\
\email{ishwar@ncra.tifr.res.in}
\and ICREA \& Institut de Ciencies de l'Espai,
Campus UAB, Facultat de Ciencies - Universitat Aut\`onoma de Barcelona
Torre C5, parell, 2da planta, Bellaterra - 08193 Barcelona, Spain \\
\email{dtorres@aliga.ieec.uab.es}
\and Instituto Argentino de Radioastronom\'{\i}a, C.C.5,
(1894) Villa Elisa, Buenos Aires, Argentina\\
\email{romero@iar.unlp.edu.ar}
\and Max Planck Institut f\"ur Kernphysik, Saupfercheckweg 1, Heidelberg 69117, Germany \\
\email{vbosch@mpi-hd.mpg.de}
\and
Grupo de Investigaci\'on FQM-322,
Universidad de Ja\'en, Campus Las Lagunillas s/n, Edif. A3, 23071 Ja\'en, Spain \\
\email{jrssutil@hotmail.com}
\and Facultad de Ciencias Astron\'omicas y Geof\'{\i}sicas, UNLP, 
Paseo del Bosque, 1900 La Plata, Argentina \\
\email{romero@fcaglp.unlp.edu.ar}
}

\date{Received; accepted }

\abstract
{A considerable fraction of the $\gamma$-ray sources discovered with 
the Energetic Gamma-Ray Experiment Telescope (EGRET) remain unidentified. 
The EGRET sources that have been properly identified are either pulsars 
or variable sources at both radio and gamma-ray wavelengths. Most of the variable sources are strong radio blazars.
However, some low galactic-latitude EGRET sources, with highly variable $\gamma$-ray emission, 
lack any evident counterpart according to the radio data available until now.}
{The primary goal of this paper is to identify and characterise the potential radio counterparts of four highly variable $\gamma$-ray sources in the galactic plane through mapping the radio surroundings of
the EGRET confidence contours and determining the variable radio sources in the field whenever possible.}
{We have carried out a radio exploration of
the fields of the selected EGRET sources using the 
Giant Metrewave Radio Telescope (GMRT)
interferometer at 21~cm wavelength, with pointings being separated by months.}
{We detected a total of 151 radio sources. Among them, we identified a few
radio sources whose flux density has apparently changed 
on timescales of months. Despite the limitations of our search, 
their possible variability makes these objects a top-priority target for multiwavelength studies of 
the potential counterparts of highly variable, unidentified gamma-ray sources.}  
{}

\keywords{
$\gamma$-ray sources -- radio sources -- microquasars -- microblazars 
}

\maketitle

\section{Introduction} 

The high-energy sky revealed by the present and past generations of $\gamma$-ray telescopes and satellites 
is populated by a large number of unidentified sources. For instance, the Third EGRET catalogue (\cite{hartman})
contains 271 entries and nearly two thirds of these $\gamma$-ray sources presently remain unidentified. 
At low galactic latitudes ($|b|\le$10$^{\circ}$), 40 of them do not show any positional
coincidence (within the 95\% EGRET contour, i.e., a size of about 0.5-1$^{\circ}$) with possible $\gamma$-ray 
objects known in our Galaxy (\cite{romero}; \cite{ torres01a}). Since it is most unlikely 
that all these sources are extragalactic, they should belong to one or more populations of galactic $\gamma$-ray
sources yet to be discovered. In particular, both observational and theoretical arguments point to the idea of galactic
sources of relativistic jets (e.g. microquasars and microblazars) being behind some of the EGRET unidentified sources. 
For instance, the high-mass X-ray binaries 
LS 5039 and LS~I+61~303 have been reported as likely 
counterparts to 3EG J1824$-$1514 and 3EG J0241+6103, respectively (\cite{k1997,paredes00}).
Such an association has been strongly supported by the detection, at TeV energies, of LS 5039 by
the High Energy Stereoscopic System (H.E.S.S., \cite{aharonian}) 
and of LS~I+61~303 by the Major Atmospheric Gamma Imaging Cherenkov telescope (MAGIC, \cite{albert}). 
In addition, theoretical models have been developed that consistently explain the high-energy gamma-ray emission 
in terms of either external and synchrotron self-Compton processes in the jets (\cite{bernado}; \cite{br2005}) 
or hadronic interactions with wind material (\cite{romero03}; \cite{orellana07}).
Alternatively, both theroretical work (\cite{dub2006}) and observational data (\cite{d2006}) 
have been used to claim that this emission may also come from a pulsar wind scenario in cases such as LS~I+61~303.
The jet or pulsar scenario remains at present a matter of interesting debate (\cite{romero07}).  

\begin{table*}
\begin{center}
\caption[]{\label{pointings} Right ascension and declination (J2000.0) for the pointing centres
of GMRT mosaics of the EGRET sources observed in this work.}
\begin{tabular}{ccccc}
\hline
\hline \noalign{\smallskip}
Pointing & 3EG J1735$-$1500        &  3EG J1746$-$1001    & 3EG J1810$-$1032   & 3EG J1904$-$1124  \\
   Id.   & (hms, $^{\circ}~^{\prime}~^{\prime\prime}$) & (hms, $^{\circ}~^{\prime}~^{\prime\prime}$) & (hms, $^{\circ}~^{\prime}~^{\prime\prime}$) & (hms, $^{\circ}~^{\prime}~^{\prime\prime}$) \\
\noalign{\smallskip} \hline \noalign{\smallskip}
0 &  17 35 52.80   $-$15 00 00.0  & 17 46 00.00   $-$10 01 48.0  & 18 10 04.80   $-$10 32 24.0  & 19 04 50.40   $-$11 25 12.0 \\
1 &  17 36 42.41   $-$14 39 12.6  & 17 46 48.69   $-$09 41 00.7  & 18 10 53.57   $-$10 11 36.7  & 19 05 39.31   $-$11 04 24.7 \\
2 &  17 37 32.19   $-$14 59 58.7  & 17 47 37.49   $-$10 01 47.1  & 18 11 42.45   $-$10 32 23.1  & 19 06 28.34   $-$11 25 11.0 \\
3 &  17 36 42.57   $-$15 20 46.7  & 17 46 48.80   $-$10 22 34.9  & 18 10 53.68   $-$10 53 10.8  & 19 05 39.43   $-$11 45 58.8 \\
4 &  17 35 03.03   $-$15 20 46.7  & 17 45 11.20   $-$10 22 34.9  & 18 09 15.92   $-$10 53 10.8  & 19 04 01.37   $-$11 45 58.8 \\
5 &  17 34 13.41   $-$14 59 58.7  & 17 44 22.51   $-$10 01 47.1  & 18 08 27.15   $-$10 32 23.1  & 19 03 12.46   $-$11 25 11.0 \\
6 &  17 35 03.19   $-$14 39 12.6  & 17 45 11.31   $-$09 41 00.7  & 18 09 16.03   $-$10 11 36.7  & 19 04 01.49   $-$11 04 24.7 \\
\noalign{\smallskip} \hline
\end{tabular}
\end{center}
\end{table*}

Here we will focus our attention on highly variable unidenti-
fied EGRET sources, defining the sample as those presenting
a variability index $I \ge 2.5$ as given by Torres et al. (2001a), which by being a relative comparison
places them more than 3$\sigma$ away from
statistical variability of pulsars. Other variability indices have
been introduced (see e.g., the $\delta$-index of \cite{nolan03}).
Although statistically correlated (\cite{torres01b}), the
specific classification of a given source can vary in each scheme, i.e., using the $I$ or $\delta$ indices. For
the four sources herein analysed, the $\delta$-index is also
compatible with they being gamma-ray variables.
Variability is naturally expected in a microquasar/microblazar scenario
due to several causes such as jet precession, motion in an eccentric orbit, and accretion
rate changes due to stellar wind inhomogeneites, although such behaviour does not exclude that we are
dealing with active galactic nuclei (AGNs),
which are also known to be variable sources of $\gamma$-rays. Models also predict that these variations
should be reflected not only in $\gamma$-rays but also in the jet non-thermal radio emission.

To determine the possible counterpart of selected EGRET sources, we undertook two campaigns
with different radio  interferometers to search for variable radio counterparts of a sample of variable unidentified EGRET sources. 
In the 2004 campaign we used the Westerbork Synthesis Radio Telescope (WSRT) for multiepoch
radio observations of three of the most variable EGRET sources at low galactic latitudes,
to determine that several radio variables were present in their location error box with their flux
density changing in more than a 30\% amplitude on timescales of months (\cite{paredes05}).
In the 2005 campaign, we conducted observations with the
GMRT of the remaining four EGRET sources.
This paper is devoted to presenting an account of these results.

\section{GMRT observations}

The observed fields, each about one square degree, were those corresponding to EGRET sources
3EG~J1735$-$1500, 3EG~J1746$-$1001, 3EG~J1810$-$1032 and 3EG J1904$-$1124.

The radio observations were carried out with the GMRT of the National Centre for Radio Astrophysics (NCRA)
in Khodad (India), during February 23 and April 19, 2005. The observations were made at the 1.4~GHz frequency
(21~cm wavelength) in spectral-line mode with 128 channels covering a 32~MHz bandwidth, with two polarizations
and two sidebands. The full-array synthesised beam of the GMRT interferometer was
about 2-3$^{\prime\prime}$ with the field-of-view limited by a $24^{\prime}$ full width half maximum (FWHM)
primary beam.
To cover the larger ($\sim1^{\circ}$) $\gamma$-ray error boxes with a sensitivity as uniform as possible,
each EGRET source field was covered with a hexagonal pattern of 7 pointings.
In this mosaicing approach, one pointing was centred on the nominal EGRET position
and the other six offset by one primary beam FWHM with position-angle increments of
$60^{\circ}$ in galactic coordinates. Their right ascension and declination are listed in Table \ref{pointings}.
We devoted an integration time of about 20 minutes to each pointing. In this way,
we were always able to map the whole solid angle
of the EGRET 68\% confidence contours and usually most of the
95\%, too. This technique was also applied when using the
Very Large Array (VLA) and the WSRT to map the fields of 3EG~J1928+1733,
3EG~J2035+4441 and 3EG~J1812$-$1316 (\cite{paredes05}).

The calibration of amplitude and bandpass was achieved by observing 3C~286 and 3C~48, whereas phase calibration 
was performed through repeated scans of the nearby phase calibrators J1733$-$130, J1822$-$096, and J1911$-$201. 
The first two fields used both interleaved scans of J1733$-$130.
Correction of the GMRT flux densities for the increase in the sky temperature in the direction of the pointings 
was also taken into account. The GMRT data was processed using standard procedures within the 
Astrophysical Image Processing System (AIPS) software 
package of NRAO. Self-calibration was possible in most of the pointings with some exceptions when no
suitable bright sources were available in the field. The mosaicing of the pointings for each field was carried
out using the FLATN task of AIPS, which includes weighting according to the
primary beam response at each pixel position.

\section{Results}

\begin{figure}
\resizebox{\hsize}{!}
{\includegraphics{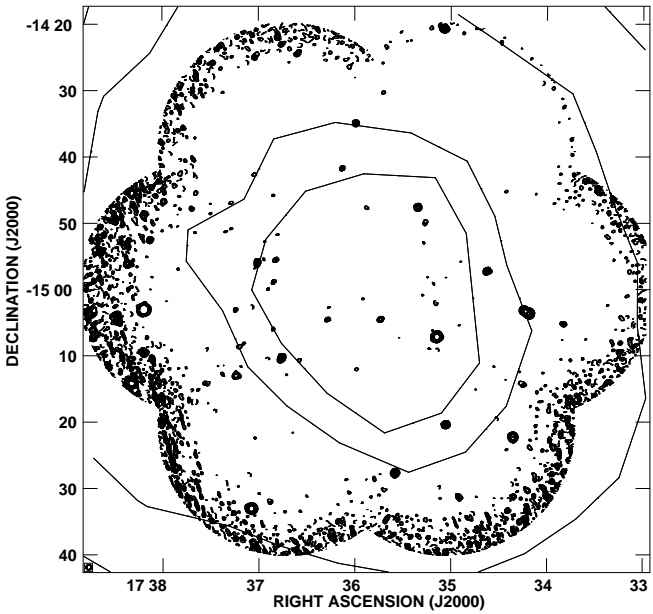}}
\caption{The 1.4 GHz image of the error box of the unidentified EGRET source
3EG~J1735$-$1500 obtained with the GMRT using a mosaicing technique.
This image has been restored with a $45^{\prime\prime}$ circular beam 
for easier display purposes. Shown contours correspond to $-5$, 5, 8, 10, 15, 20, 30, 40, 50 and 100
times 0.4 mJy beam$^{-1}$.
The noisy appearance toward the edges is due to strong primary beam correction.
The 50\%, 68\%, 95\%, and 99\% confidence contours for
the location of the EGRET source are also shown. Forty-three radio sources are clearly detected in this 
field, but only two of them show any evidence of variability. 
}
\label{1735}
\end{figure}

We show in Fig.~\ref{1735} an example of one of the obtained radio maps, which corresponds to a GMRT mosaic of the source 3EG~J1735$-$1500.
It was computed using uniform weight. We inspected each 
of the error boxes of the four EGRET sources observed and searched for all radio sources in the field.
This was achieved through a combined use of visual inspection for those obvious cases and automated source extraction procedures,
such as the {\it Search and Destroy} (SAD) 
task contained in the AIPS package. Only objects with peak flux densities higher than 4-5 times
the root mean square (rms) noise were retained.
For each radio source detected, we measured its peak ($S_{\nu}^{\rm Peak}$) and the total flux density 
($S_{\nu}^{\rm Integ}$), position and angular size by means of an elliptical Gaussian fit.
This procedure was carried out separately for the February 23 
(first epoch) data, the April 19 (second epoch) data, and for the combined maps from both epochs.
At the end of this process, all source detections were finally revised manually to ensure their reality.
Artifacts near bright sources that could be considered low-significance detections by SAD were removed
and the fitting of close double radio sources was also checked individually.

\subsection{Catalogue}

Based on the combined maps from the two observing epochs,
we detected a total of 151 radio sources above
a flux density of $\sim$1.5 mJy. 
Typical rms noises achieved in each pointing are in the 0.1-0.4 mJy range depending on particular
observing conditions and bright sources limiting the dynamical range.
The strongest source detected is GMRT~J173811.6$-$150301,
with an average peak flux density of 460.8 mJy beam$^{-1}$.

In Table~\ref{list}, available in electronic form 
in the online material of this paper, we list all the radio sources detected. From the 1st to 8th columns, this table contains 
the GMRT name for each entry, the J2000.0 coordinates, the peak flux density, 
the integrated flux density and the apparent angular size. The corresponding values were determined by fitting elliptical
Gaussians using the AIPS task JMFIT. Uncertainties quoted in 4th and 5th columns
are based on the formal errors of the fit and allow the reliability of the detection to be judged. However, they do not include the
contribution of primary beam corrections as a function of angular distance $\theta$ to the phase
centre (9th column). An estimate of the combined error is given in the 10th and 11th columns for both the peak
and integrated flux density, respectively. The final 12th column contains a radio variability index expressing the flux density difference
between the two observing epochs in terms of this sort of combined noise estimator (see below and Appendix A for details.) This information
has been omitted for a few cases where an unreliable result is suspected. This is usually connected with some extended or faint sources
and occasional Gaussian fitting problems.

The GMRT positions are usually accurate to better than one arc-second,
which is suitable for identifying optical/near infrared counterparts in follow-up observations
even in relatively crowded fields. A preliminary search has been conducted in some cases
by inspecting the plates from the Digitized Sky Survey (DSS, \cite{lasker})
and the 2 Micron All Sky Survey (2MASS, \cite{st2006}) as discussed below.

\subsection{Search for variable radio sources}

Based on previous radio work, such as the 
the GT galactic plane patrol  (\cite{gt}) or the FIRST survey (\cite{vries}),
one could expect that 1\% to 5\% of the sources we detected could be intrinsically variable. This would 
translate into one or two peculiar variable objects expected in this work.

The search for variables in the GMRT data was first carried out systematically by
plotting the flux densities of sources detected at two different epochs (February and April), one versus the other. 
Of course, two epochs of observation are not sufficient to clearly establish an object's variability unless
the amplitude of variation is rather high. However, this is the kind of object that we are looking for in this work
as radio counterparts. In this context, a threshold of variability amplitude
of $\pm30$\%, which is a substantial fraction of the total flux, appeared reasonable for our purposes. 

\begin{figure}
\resizebox{\hsize}{!}
{\includegraphics[angle=-90]{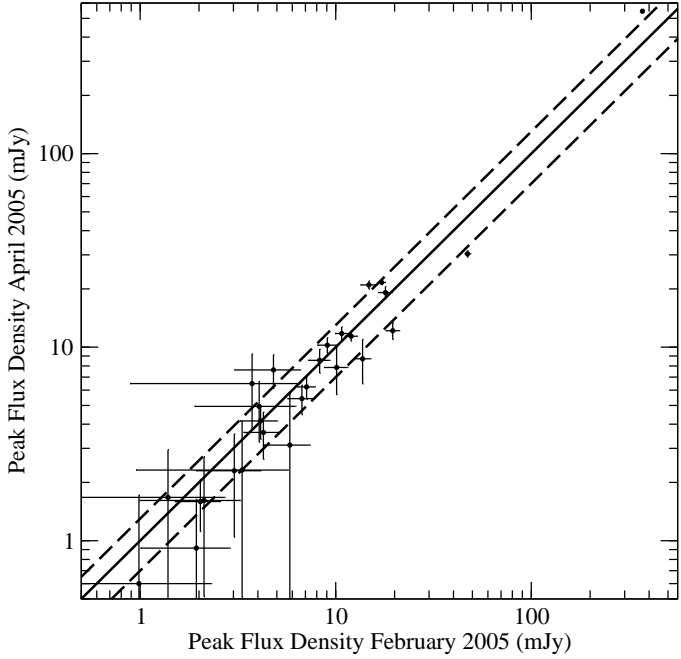}}
\caption{Flux density in
February 2005 versus flux density in April 2005 for all
compact radio sources in the field of 3EG~J1735$-$1500
detected with the GMRT at 21 cm. Error bars shown
are $\pm3$ times the JMFIT rms. The dashed lines represent a variability amplitude
of $\pm30\%$ above which an object is considered
a likely radio counterpart of the unidentified variable
EGRET source. Two radio sources in this field are
found to satisfy such criterion but their large angular distance from the phase
centre makes their variability suspicious due to uncertain primary beam correction.} 
\label{compa}
\end{figure}

One example of a variability plot is illustrated in Fig.~\ref{compa}
in the case of 3EG~J1735$-$1500 based on our two epoch GMRT observation separated by a few months. 
Here, two radio sources in the field stand out as apparent radio variables. 
In the four EGRET fields observed, a total of 11
candidate variables were initially selected in this way.
However, as quoted above, the absolute flux densities of our sources could be additionally affected by uncertainties 
in the primary beam correction applied to them.
The net accuracy of pointing and tracking of GMRT antennas is about 2 to 3 arcmin, which at L-band corresponds to about 10\% 
of the FWHM. This leads to poor primary beam correction resulting in significant uncertainties in the absolute flux densities 
of sources away from the pointing centre (see Appendix A). 
In view of this problem, we instead define a radio variability index measuring the significance of 
this difference between a source's individual peak flux densities from its average value $\bar{S}_{\nu}^{\rm Peak}$ as
\begin{equation}
{\rm Var.~Index} = \sqrt{ \left[\frac{S_{\nu}^{\rm Peak,1} - \bar{S}_{\nu}^{\rm Peak}}{{\rm Err}^{\rm Peak,1}}\right]^2 
                       +  \left[\frac{S_{\nu}^{\rm Peak,2} - \bar{S}_{\nu}^{\rm Peak}}{{\rm Err}^{\rm Peak,2}}\right]^2
  }, 
\end{equation}
where the combined JMFIT+primary beam correction error is computed as 
\begin{equation}
{\rm Err^{{\rm Peak},i}(\theta)} = \sqrt{ {\rm [rms~JMFIT]}^2 +   \left[S_{\nu}^{{\rm Peak},i} \frac{\Delta P_b}{P_b(\theta)^2}\right]^2 }
\end{equation}
for each of the two observing epochs ($i=1,2$). Here, the second term represents the error when dividing by the primary beam
response $P_b(\theta)$ at the source location (Eq. \ref{errorbeam}). 
If a source has a variability index above say 3,
we consider it as a candidate variable radio source.

Only two radio sources in our catalogue turned out 
to have a reliable variability index approaching 3 or higher, assuming a typical GMRT pointing error of $\Delta \theta = 2^{\prime}$.
The observed parameters of these candidate variables are
separately listed in Table~\ref{taula}, bearing in mind that variability still needs to be confirmed.
Notation here is similar to Table~\ref{list}.
The variability indices for the rest of the GMRT radio sources are consistent with being non-variable.
Of course, the possible detection of variability does not ensure an immediate connection with the corresponding EGRET source.
Indeed, many extragalactic radio sources, such as quasars and other AGNs, are known to be variable in the radio.
Despite the limitations of our variability analysis, the candidate variables 
reported here do represent good targets for follow-up observations that could
ultimately reveal clues leading to the final identification of the EGRET source, 
especially if their position is confirmed with a better precision by future $\gamma$-ray telescopes 
such as the Gamma-ray Large Area Space Telescope (GLAST).

\begin{table*}
\begin{center}
\caption[]{\label{taula} Candidate variable radio sources detected with the GMRT inside the error box of 
one of our selected $\gamma$-ray  variable EGRET sources.}
\begin{tabular}{cccccrcrcc}
\hline
\hline \noalign{\smallskip}
Source  & GMRT Id. & $\alpha_{\rm J2000.0}$   & $\delta_{\rm J2000.0}$  & $\theta$ & $S_{\nu}^{{\rm Peak},1}$ & Err$^{{\rm Peak},1}$ &  $S_{\nu}^{{\rm Peak},2}$ & Err$^{{\rm Peak},2}$ & Var. \\
3EG~J   &         & (hms)               & ($^{\circ}~\arcmin~\arcsec$)    & ($^{\prime}$) &  (mJy/b)  & (mJy/b) &  (mJy/b) & mJy/b & Index             \\
\noalign{\smallskip} \hline \noalign{\smallskip}
1904$-$1124  &  J190601.7$-$112510  &  19 06 01.781(0.009)   &  $-$11 25 10.10(0.20)   & 6.5 &   5.1 $\pm$ 0.3 & 0.8 &    9.1 $\pm$ 0.6 & 1.5 & 3 \\
             &  J190617.4$-$112850  &  19 06 17.428(0.003)   &  $-$11 28 50.11(0.04)   & 4.5 &  16.2 $\pm$ 0.3 & 1.5 &   26.1 $\pm$ 0.5 & 2.4 & 4 \\
\noalign{\smallskip} \hline
\end{tabular}
\end{center}
\end{table*}

\subsection{Resolved and multiple radio sources} \label{mult}
 
\begin{figure}
\resizebox{\hsize}{!}
{\includegraphics[angle=-90]{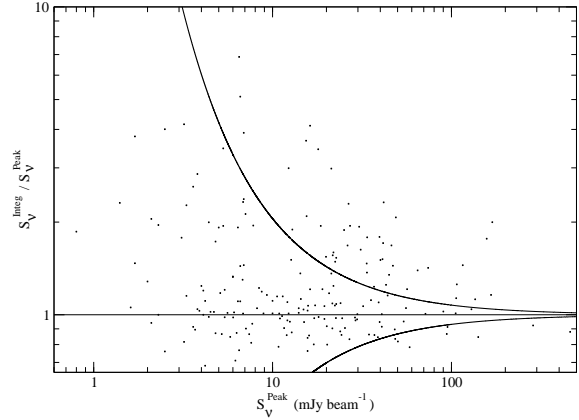}}
\caption{Plot of the integrated to peak flux density ratio
for all detections in this work as a function of peak flux density.
Continuous lines represent the upper and lower envelopes of the plot region considered
to correspond to unresolved radio sources according to an approach similar
to the one in Bondi et al. (2007). Objects located above the upper envelope line
appear to be resolved by our GMRT observations.}
\label{bondi}
\end{figure}

We also looked for sources with a resolved structure. For each detection
we compared the peak 
$S_{\nu}^{\rm Peak}$ with the integrated flux density $S_{\nu}^{\rm Integ}$ to differentiate among resolved and 
unresolved sources.
The maps from the two epochs combined were used for this purpose.
We plotted in Fig.~\ref{bondi} the ratio $S_{\nu}^{\rm Integ}$/$S_{\nu}^{\rm Peak}$  
versus $S_{\nu}^{\rm Peak}$. Following Bondi et al. (2007), we roughly
estimated the lower envelope of the 
data shown in Fig.~\ref{bondi} by fitting the equation 
\begin{equation}
S_{\nu}^{\rm Integ}/S_{\nu}^{\rm Peak} = a^{-(b/S_{\nu}^{\rm Peak})}  \label{envel}
\end{equation}
where $a=0.75$ and $b=-0.25$. This curve was later mirrored above with respect to the
$S_{\nu}^{\rm Integ}$/$S_{\nu}^{\rm Peak}$ = 1 line.
Data points in between the lower and upper mirrored envelopes are believed to correspond to point-like
sources and to their dispersion due to statistical errors. In contrast, points lying above the upper envelope
are considered as resolved sources. They represent about 23\% out of the total detected sources,
although this value must be taken as an upper limit because some of them are grouped as double sources.
Double or triple morphologies are often found among them and
none exhibited structural changes between the two epochs of observation.

\begin{figure}
\resizebox{\hsize}{!}
{\includegraphics[angle=-90]{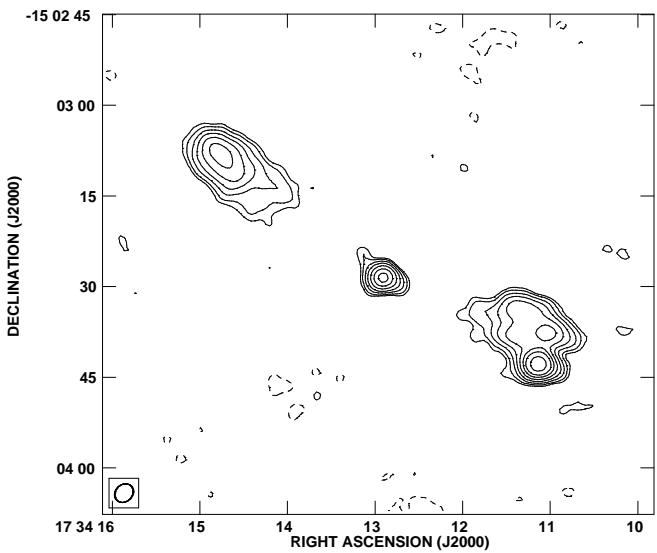}}
\caption{Triple radio source in the field of the unidentified EGRET source 3EG~J1735$-$1500 obtained 
after combining the February and April runs with the GMRT. 
The contours shown are $-$3, 3, 5, 9, 
15, 25, 40, 60, 100 and 150 times the rms noise of 0.13 mJy beam$^{-1}$. 
The corresponding synthesised beam is shown in the
bottom left corner, and it corresponds to 3\pri 46$\times$2\pri 91, with a position angle of $-$36\grp 8.}
\label{figtriple}
\end{figure}

\begin{table*}
\begin{center}
\caption[]{\label{triple} The triple radio source within the error box of 3EG~J1735$-$1500.}
\begin{tabular}{ccccc}
\hline
\hline \noalign{\smallskip}
 Component & $\alpha_{\rm J2000.0}$   & $\delta_{\rm J2000.0}$  & $S_{\nu}^{\rm Peak}$   &   $S_{\nu}^{\rm Integ}$   \\
           & (h m s)               & ($^{\circ}~\arcmin~\arcsec$)    & (mJy beam$^{-1}$)  &      (mJy)             \\
\noalign{\smallskip} \hline \noalign{\smallskip}

West lobe$^{(*)}$  &   17 34 11.126(0.002)  &  $-$15 03 42.65(0.03)  &   21.9 $\pm$ 0.4  &     27.1  $\pm$ 0.7  \\
 Core              &   17 34 12.899(0.003)  &  $-$15 03 28.55(0.05)  &   12.5 $\pm$ 0.4  &     11.8  $\pm$ 0.6  \\
East lobe          &   17 34 14.720(0.020)  &  $-$15 03 09.00(0.20)  &    6.6 $\pm$ 0.4  &     34.0  $\pm$ 2.0   \\
\noalign{\smallskip} \hline
\end{tabular}
~\\
(*) This lobe is decomposed into two source components in Table \ref{list}.
\end{center}
\end{table*}

\section{Discussion: individual EGRET fields}

\subsection{3EG~J1735$-$1500}

This is the most variable EGRET source in our sample ($I=8.86$, \cite{torres01a}, see, however, \cite{nolan03}).
Inside its EGRET error box, Mattox et al. (2001) reported the presence of 
PMN J1738$-$1502, a source from the Parkes-MIT-NRAO survey (PMN, \cite{gr1994}),
as a potential radio counterpart with a low a priori probability that this object
is a $\gamma$-ray blazar. In an analysis carried out later by Sowards-Emmerd et al. (2004), 
they classify this radio source as a high-confidence blazar. PMN J1738$-$1502 appears to be coincident with our
variable source GMRT J173811.6$-$150301 in Table \ref{taula}, with the observed variability in agreement
with the proposed blazar nature. 
Combi et al. (2003) confirmed as well the presence of PMN J1738$-$1502 as a flat-spectrum compact radio source,
with a flux density of 330 mJy at 1.4 GHz. In addition, by using the NRAO VLA Sky Survey
(NVSS; \cite{condon}) they found a total of 23 radio sources with flux density greater
than 10 mJy at 1.4 GHz within the inner 95$\%$ confidence contour of 3EG~J1735$-$1500.

In the Combi et al. (2003) exploration of the 3EG~J1735$-$1500 error box, they also
reported the presence a new radio galaxy (a double-sided source of Fanaroff-Riley type II) inside the 95\% EGRET confidence contour.
The radio galaxy's central core has an estimated  
position of 
$\alpha_{\rm J2000.0}$=17h37m12.9s$\pm$0.3s, 
$\delta_{\rm J2000.0}$=$-15^{\circ}11^{\prime}0.20^{\prime\prime}\pm15^{\prime\prime}$, but nothing is detected
here in our GMRT maps. Actually, this is not totally unexpected since this position 
lies very close to mosaic-pointing edges where the primary beam correction increases
the noise significantly. Based on the NVSS peak flux density, one would expect a source at
merely 4-5 sigma level and; therefore, its non detection does not come as a surprise.
On the other hand, removing the shortest GMRT baselines to enhance compact sources
makes the radiogalaxy extended lobes fully resolved in our GMRT maps.

Our observations have revealed a total of 43 sources in this field, all listed in Table~\ref{list}. 
Four of them have been resolved by the GMRT interferometer presenting either
a double structure (GMRT~J173345.7$-$151643, GMRT~J173421.2$-$152222, and GMRT~J173704.3$-$153301)
or even a triple structure (GMRT~J173412.8$-$150328).
We show in Fig.~\ref{figtriple} the contour map of this last interesting triple object, obtained after combining 
the February and April runs. The position of the components and their peak and integrated flux 
densities are given in Table~\ref{triple}.
The morphology of this source thus appears clearly reminiscent of a Fanaroff-Riley type
II radio galaxy, with a core and two lobe components.
One of the lobes has a clear and bright hot spot.
The three components are well-aligned in the northeast-southwest direction, 
with the outer components separated from the core by an angular distance of about $30^{\prime\prime}$. 

No reliable radio variables have been detected in this field.
Bosch-Ramon et al. (2006a)
applied a microquasar model to explain the high-energy $\gamma$-ray emission 
of 3EG~J1735$-$1500 consistent with the observations at lower energies
(from radio frequencies to soft $\gamma$-rays) within the EGRET error box. Although their theoretical model suggests
that a microquasar might be the counterpart of this particular source, 
other alternatives cannot be ruled out as possible counterparts (e.g., \cite{punsly}, \cite{br2006b}).

\subsection{3EG~J1746$-$1001}

We have detected 36 sources within the error box of this EGRET source. 
Two of them were reported by Mattox et al. (2001) as potential radio counterparts 
of 3EG~J1746$-$1001 with a low a priori probability. These sources are PMN J1744$-$1011 and PMN J1747$-$0959 
with a flux density at 5 GHz of 85 and 61 mJy, respectively. We have also detected these sources at 
1.4 GHz showing a double structure. 

For the first one PMN J1744$-$1011, we have resolved an elongated 
northeast-southwest structure with two strong components. GMRT~J174443.2$-$101001 is 
the northern component, with a peak flux density of 46.2$\pm$0.7 mJy beam$^{-1}$ and an integrated  
flux density of 77.6$\pm$1.4 mJy. GMRT~J174441.9$-$101040 is the southern component,
which appears with a peak flux density of 106.3$\pm$0.7 mJy beam$^{-1}$ and an integrated  
flux density of 120.3$\pm$1.1 mJy. 
For the second one PMN J1747$-$0959, our GMRT map shows a double source elongated in the east-west direction. 
The east component, GMRT~J174727.8$-$095917, has a peak flux density of 129.4$\pm$0.4 mJy beam$^{-1}$ and 
an integrated  flux density of 134.8$\pm$0.6 mJy. The peak and the integrated flux density of the western 
component, GMRT~J174727.2$-$095911, amount to 12.1$\pm$0.4 mJy beam$^{-1}$ and 13.3$\pm$0.6 mJy, respectively. 

We also found other five sources in this field showing extended or double structures, 
namely GMRT~J174457.8$-$101206,  GMRT~J174501.4$-$093849, GMRT~J174556.0$-$100613, GMRT~J174616.5$-$102358, and GMRT~J174624.1$-$095208. 
Information about their peak and integrated flux densities, apparent size,  
and position angle of their components can be found in Table~\ref{list}.

No reliable variables were found in the field of 3EG~J1746$-$1001. An interesting object is, however,
GMRT J174535.5$-$101439, which
has a barely resolved core+one-sided jet morphology.
No obvious counterpart is present at optical, infrared, or X-ray according to the inspected
surveys.

\subsection{3EG~J1810$-$1032}

We detected 38 sources within the error box of this EGRET source. For this $\gamma$-ray source, 
three different potential radio counterparts with low a priori probability were proposed by Mattox et al. (2001). 
These sources were PMN J1808$-$1041, PMN J1810$-$1054, and PMN J1810$-$1102 
with a flux density at 5 GHz from single-dish surveys of 48, 49, and 103 mJy, respectively.
We have not clearly detected any of them and this is likely due to lack of sensitivity to very
extended objects more easily detected in single dish surveys. 

Among the 38 sources detected, there are five of them that have been resolved with the 
GMRT showing a double or marginally resolved double structure. These sources are GMRT~J180809.1$-$104031, 
GMRT~J180834.3$-$103024, GMRT~J180943.4$-$104055, GMRT~J181017.7$-$102907, and GMRT~J181030.9$-$101839.

No reliable radio variables were found in this field.

\subsection{3EG~J1904$-$1124}

We detected 34 sources within the error box of this EGRET source. 
Two different potential radio counterparts,  with a low a priori probability,
were proposed by Mattox et al. (2001). These sources were PMN J1905$-$1153 and PMN J1906$-$1114 
with a flux density at 5 GHz of 197 and 126 mJy, respectively. 

The first of them has been classified 
as a plausible blazar by Sowards-Emmerd et al. (2004). We also detected this source, 
with a peak flux density of 286.2$\pm$1.4 mJy beam$^{-1}$ 
and an integrated flux density of 264.2$\pm$2.0 mJy. These peak and integrated flux densities fall slightly offset from 
the envelope curve quoted in Sect. \ref{mult}, but yet they can be considered
to be consistent with an unresolved source. 
Although our data at 1.4 GHz and the 5 GHz data were not taken simultaneously, 
the results seem to point to a non-thermal emission. 
The second source, PMN J1906$-$1114, also detected by us, shows an elongated structure with two components. 
The strongest one is GMRT~J190645.1$-$111434, with a peak flux density of 115.7$\pm$0.7 mJy beam$^{-1}$ 
and an integrated flux density of 167.6$\pm$1.3 mJy. The fainter component is GMRT~J190644.8$-$111416 with a 
peak and an integrated flux density of 33.6$\pm$0.7 mJy beam$^{-1}$ and 49.3$\pm$1.3 mJy, respectively. 
In this source, the northern component could have a non-thermal nature, whereas 
it is not as clear for the southern component.
In the error box of 3EG~J1904$-$1124, we also detected four other extended sources, namely
GMRT~J190339.9$-$114756, GMRT~J190341.1$-$112310,   
GMRT~J190430.2$-$115241 and GMRT~J190523.0$-$110250.

In the field of 3EG~J1904$-$1124, we found two sources that are candidate radio variables. Their
expected flux density uncertainty due to primary beam correction is at the $\sim10$\% level.
Both of them have a compact appearance.

\section{Conclusions}

We have reported the radio results of source detection and of the search for variables positionally consistent with the
highly variable, unidentified EGRET sources, at the 21 cm wavelength, and using a mosaicing technique
with GMRT. 
The targets studied in this work include 
3EG~J1735$-$1500, 3EG~J1746$-$1001, 3EG~J1810$-$1032 and 3EG J1904$-$1124.
Our main findings can be summarised as follows,

\begin{enumerate}

\item The number of confident detections within the error box of each EGRET source
quoted above was 43, 36, 38, and 34 objects, respectively. Our limiting flux density is not
uniform across all pointings but typically amounts to $\sim 1.5$ mJy.
The mapped field of view always covered the full solid
angle of the 68\% EGRET confidence contours and usually most of the 95\% as well. 

\item Out of the 151 sources detected in total, only 
a few of them displayed apparent radio variability with reliable amplitude
on a two-month time baseline in two of the EGRET fields.
Although a variability search was the original motivation for this work, we have been severely limited
by instrumental problems in this task, and not all the detected sources could be suitably explored in their time behaviour. Nevertheless, 
we have undertaken a programme for future follow-up observation of the proposed candidate variables, 
aimed in particular at finding their near-infrared and X-ray counterparts. 
Although we anticipate here that most
will turn out to be unrelated background sources, strong attention should be given to any of them found to be positionally
coincident with the $\gamma$-ray emission detected by future space
observatories such as
GLAST with improved source localization

\item As a byproduct of this work, we have identified 16 double radio sources, 4 marginally resolved double sources, and one interesting
triple radio source positionally consistent with 3EG~J1735$-$1500. Its morphology is 
remarkably reminiscent of a Fanaroff-Riley type II radio galaxy. However, the absence of an optical/near infrared
counterpart for the central core at present precludes assessing its true nature and connection with
the EGRET emission.

\end{enumerate}

\begin{acknowledgements}
The authors acknowledge support by the Ministerio de
Educaci\'on y Ciencia (Spain) under grants AYA2007-68034-C03-01, AYA2007-68034-C03-02, AYA2006-00530, 
FEDER funds and Plan Andaluz de Investigaci\'on, Desarrollo e Innovaci\'on (PAIDI)
of Junta de Andaluc\'{\i}a as research group FQM322. D.F.T. also acknowledges partial support from  the
Guggenheim Foundation. G.E.R is
supported by the Argentine Agencies CONICET (PIP 5375) and ANPCyT (PICT 03-13291). J.A.C. is a researcher of the programme {\em Ram\'on y
Cajal} funded jointly by the Spanish Ministerio de Educaci\'on y Ciencia (former
Ministerio de Ciencia y Tecnolog\'{\i}a) and Universidad de Ja\'en.
During this work, V.B-R. has been supported by the DGI of the Ministerio de
Ciencia y Tecnolog\'{\i}a (Spain) under the fellowship FP-2001-2699. V.B-R. thanks the Max-Planck-Institut 
f\"ur Kernphysik for its support and kind hospitality. V.B-R. gratefully 
acknowledges support from the Alexander von Humboldt Foundation. GMRT is run by the National Centre for Radio Astrophysics of the Tata Institute of Fundamental Research.
  
\end{acknowledgements}

\Online

\begin{appendix}

\section{Uncertainty in absolute flux density due to pointing/tracking offsets}

\begin{figure}
\resizebox{\hsize}{!}
{\includegraphics{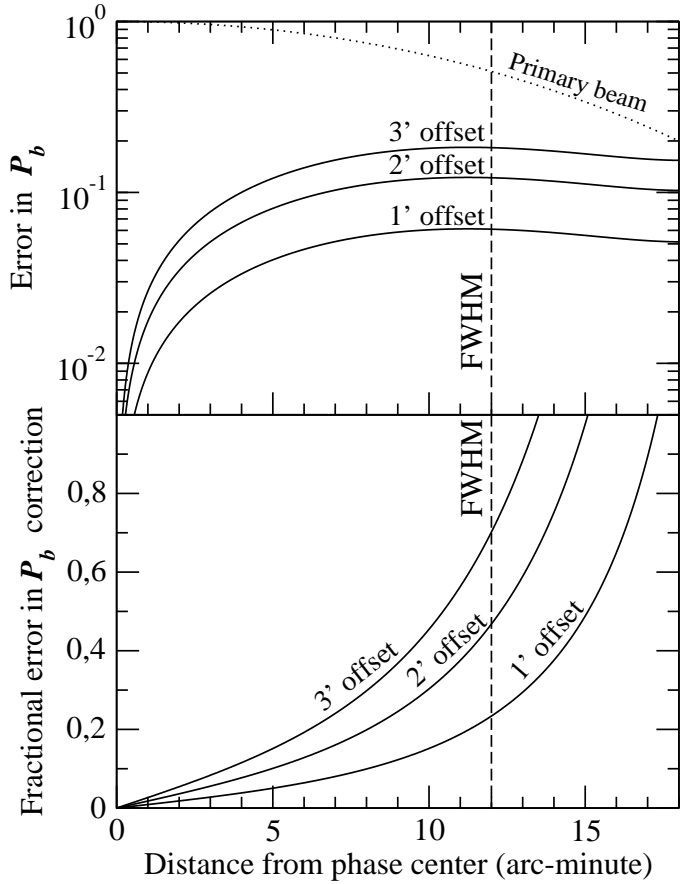}}
\caption{{\bf Top.} Uncertainty in the GMRT primary beam response at the 21 cm wavelength for different
values of the pointing/tracking offset as a function of the distance from the phase centre.
The shape of the primary beam is also plotted as a pointed line and the vertical dashed line indicates its FWHM.
{\bf Bottom.} Fractional error introduced when the synthesised map is divided by the sky position dependent
primary beam response shown in the top panel.}
\label{pberror}
\end{figure}

The properties of the GMRT antennae are such that pointing/tracking offsets 
of a few arc-minutes are not uncommon.
Consequently, the true primary beam pattern can experience a shift in this order instead of being centred exactly
at the assumed phase centre. This effect is negligible at most of the typical GMRT long wavelengths,
but not completely at the shortest one of 21 cm used here where it can reach about 10\% of the primary beam FWHM.

How does this translate into primary beam correction?
The anntena primary beam response $P_b(\theta)$ as a function of the distance $\theta$ from the phase centre
can be computed using polynomic coefficients from the GMRT User's Manual. In general, it can be expressed as
\begin{equation}
P_b(\theta)= 1 + A x^2 + B x^4 + C x^6 + D x^8, \label{eq1}
\end{equation}
where $A=-2.27961 \times 10^{-3}$, $B=21.4611 \times 10^{-7}$, $C=-9.7929 \times 10^{-10}$, $D=1.80153\times 10^{-13}$, and
$x = \nu \theta$ is the product of the observing frequency in GHz times the angular distance in arc-minutes. 
Any pixel in a synthesised radio map will have its distance to the phase centre known within an uncertainty
$\Delta \theta$. We assume here that this value is comparable to the typical GMRT pointing/tracking offset. Therefore,
the corresponding uncertainty in the primary beam response will propagate by incrementing Eq. \ref{eq1} as
\begin{equation}
\Delta P_b(\theta) = 2x(  A   + 2 B x^2  + 3 C x^4  +4 D x^6 ) \Delta x, \label{eq2}
\end{equation} 
where $\Delta x = \nu \Delta \theta$, since we assume that the frequency is known exactly.
Hereafter, we adopt $\nu=1.4$ GHz as the corresponding value for the L-band wavelength of 21 cm.

To quantify the problem, the top panel in Fig. \ref{pberror}
shows the error in primary beam response $\Delta P_b(\theta)$ computed using Eq. \ref{eq2}
as a function of $\theta$ for different offset values. The bottom panel illustrates the
corresponding fractional error,
\begin{equation}
\frac{\Delta S_{\nu}}{S_{\nu}} = \frac{\Delta P_b}{P_b^2},  \label{errorbeam}
\end{equation}
when dividing a measured flux density $S_{\nu}$, or a map pixel value,
by the primary beam response. From these plots we see that primary beam corrected flux densities may
be affected by an additional uncertainty 
as large as $\pm10$\% for objects beyond $\sim 6^{\prime}$ from the phase centre.

\end{appendix}

{\scriptsize

\longtab{4}{
\begin{longtable}{lccrrrrrrrrr}
\caption[]{\label{list} List of GMRT sources detected}\\
\hline\hline
GMRT Identification  &  ~~~~~$\alpha_{\rm J2000.0}$  &  ~~~~~~~$\delta_{\rm J2000.0}$  &  $S_{\nu}^{\rm Peak}$  &  $S_{\nu}^{\rm Integ}$  &
    $a$~~    &    $b$~~    &   P.A. & $\theta$   & Err$^{\rm Peak}$ & Err$^{\rm Integ}$ & Var. \\ 
     &          ~~~~~~(hms)           &          ~~~~~~~~($^{\circ}~^{\prime}~^{\prime\prime}$)           &    (mJy/b)   &        (mJy)            &  ($^{\prime\prime}$) &  ($^{\prime\prime}$) &  ($^{\circ}$)~ & ($^{\prime}$)   & (mJy/b) & (mJy)  &  Index   \\
\hline
\endfirsthead
\caption{continued.}\\
\hline\hline
GMRT Identification  &  ~~~~~$\alpha_{\rm J2000.0}$  &  ~~~~~~~$\delta_{\rm J2000.0}$  &  $S_{\nu}^{\rm Peak}$  &  $S_{\nu}^{\rm Integ}$  &
    $a$~~    &    $b$~~    &   P.A. & $\theta$   & Err$^{\rm Peak}$ & Err$^{\rm Integ}$ & Var. \\
     &          ~~~~~~(hms)           &          ~~~~~~~~($^{\circ}~^{\prime}~^{\prime\prime}$)           &    (mJy/b)   &        (mJy)            &  ($^{\prime\prime}$) &  ($^{\prime\prime}$) &  ($^{\circ}$)~ & ($^{\prime}$)   & (mJy/b) & (mJy)  & Index   \\
\hline
\endhead
\hline
\endfoot
J173336.1$-$150727              &  17 33 36.180(0.020) &  $-$15 07 27.40(0.20)  &   7.1(0.7)  &   6.4(0.9)  &  6.38  &  3.06  &  44.4 &   11.7 &     3.2 &     2.9 &     0.9  \\
J173345.7$-$151643$^{\rm (D)}$  &  17 33 45.793(0.003) &  $-$15 16 43.41(0.08)  &  44.0(0.9)  &  51.0(2.0)  & 10.07  &  3.45  &  23.6 &   18.0 &   114.1 &   132.3 &     0.2  \\
J173346.3$-$151627$^{\rm (D)}$  &  17 33 46.332(0.004) &  $-$15 16 27.90(0.10)  &  32.1(0.9)  &  35.0(2.0)  &  9.14  &  3.46  &  25.9 &   17.7 &    74.6 &    81.3 &     0.1  \\
J173349.7$-$150501              &  17 33 49.710(0.030) &  $-$15 05 01.90(0.40)  &   3.8(0.5)  &  11.0(2.0)  &  7.18  &  6.41  & 165.7 &    7.6 &     0.9 &     2.8 &     0.8  \\
J173411.1$-$150337$^{\rm (*)}$  &  17 34 11.110(0.020) &  $-$15 03 37.80(0.20)  &   6.5(0.4)  &  45.0(3.0)  & 11.77  &  6.74  &  57.4 &    3.7 &     0.6 &     4.4 &     1.8  \\
J173411.1$-$150342$^{\rm (T)}$  &  17 34 11.126(0.002) &  $-$15 03 42.65(0.03)  &  21.9(0.4)  &  27.1(0.7)  &  4.24  &  3.39  & 160.4 &    3.8 &     1.6 &     2.1 &     $-$  \\ 
J173412.8$-$150328$^{\rm (T)}$  &  17 34 12.899(0.003) &  $-$15 03 28.55(0.05)  &  12.5(0.4)  &  11.8(0.6)  &  3.46  &  3.11  & 143.3 &    3.5 &     0.9 &     1.0 &     0.9  \\
J173414.7$-$150309$^{\rm (T)}$  &  17 34 14.720(0.020) &  $-$15 03 09.00(0.20)  &   6.6(0.4)  &  34.0(2.0)  & 10.05  &  5.63  &  45.4 &    3.2 &     0.6 &     2.8 &     1.3  \\
J173421.2$-$152222$^{\rm (D)}$  &  17 34 21.260(0.002) &  $-$15 22 22.88(0.03)  &  38.8(0.6)  &  32.8(0.7)  &  5.30  &  3.60  &  49.7 &   10.2 &    12.3 &    10.4 &     1.0  \\
J173421.3$-$152209$^{\rm (D)}$  &  17 34 21.303(0.005) &  $-$15 22 09.56(0.06)  &  17.0(0.6)  &  13.6(0.7)  &  5.25  &  3.44  &  50.5 &   10.2 &     5.4 &     4.3 &     1.1  \\
J173427.0$-$144441              &  17 34 27.030(0.030) &  $-$14 44 41.40(0.30)  &   2.0(0.3)  &   2.6(0.5)  &  7.75  &  5.31  & 109.5 &   10.3 &     0.7 &     1.0 &     1.5  \\
J173438.3$-$145723              &  17 34 38.338(0.007) &  $-$14 57 23.84(0.09)  &   8.7(0.5)  &  10.4(0.8)  &  4.31  &  4.04  &  93.4 &    6.6 &     1.4 &     1.7 &     0.2  \\
J173439.8$-$145028              &  17 34 39.880(0.020) &  $-$14 50 28.90(0.30)  &   3.6(0.3)  &   3.3(0.4)  &  6.72  &  4.66  &  16.4 &   11.4 &     1.5 &     1.4 &     1.0  \\
J173459.2$-$142327              &  17 34 59.226(0.008) &  $-$14 23 27.70(0.20)  &   6.0(0.3)  &   8.0(0.5)  &  7.91  &  5.89  &   6.2 &   15.8 &     7.3 &     9.7 &     0.9  \\
J173501.7$-$143643              &  17 35 01.778(0.009) &  $-$14 36 43.60(0.20)  &   2.1(0.2)  &   1.9(0.3)  &  5.14  &  3.59  & 155.8 &    2.5 &     0.2 &     0.3 &     1.7  \\
J173503.5$-$152026              &  17 35 03.556(0.003) &  $-$15 20 26.73(0.05)  &  12.7(0.3)  &  19.1(0.7)  &  5.10  &  3.70  &  17.1 &    0.4 &     0.3 &     0.7 &     1.7  \\
J173504.3$-$142037              &  17 35 04.347(0.002) &  $-$14 20 37.39(0.04)  &  28.9(0.4)  &  28.3(0.5)  &  7.68  &  5.16  & 178.5 &   18.6 &    94.4 &    92.5 &     0.1  \\
J173508.7$-$150708              &  17 35 08.724(0.002) &  $-$15 07 08.13(0.01)  & 105.6(0.5)  & 106.7(0.7)  &  8.46  &  4.72  &  74.8 &   12.8 &    59.5 &    60.2 &     0.4  \\
J173520.6$-$144739              &  17 35 20.673(0.002) &  $-$14 47 39.69(0.03)  &  20.1(0.3)  &  18.4(0.4)  &  6.35  &  3.62  & 156.5 &    9.4 &     5.4 &     5.0 &     0.6  \\
J173534.9$-$152740              &  17 35 34.921(0.007) &  $-$15 27 40.75(0.07)  &  18.6(0.5)  &  36.0(2.0)  &  8.89  &  4.91  & 116.8 &   10.3 &     6.1 &    11.9 &     0.8  \\
J173536.8$-$142543              &  17 35 36.860(0.020) &  $-$14 25 43.60(0.20)  &   6.0(0.4)  &   8.0(0.6)  &  8.77  &  6.25  &  31.5 &   15.8 &     7.3 &     9.7 &     0.8  \\
J173542.0$-$143022              &  17 35 42.008(0.009) &  $-$14 30 22.70(0.20)  &   6.6(0.3)  &   7.3(0.5)  &  6.99  &  5.85  &  81.0 &   12.9 &     3.8 &     4.2 &     0.9  \\
J173543.9$-$150428              &  17 35 43.920(0.010) &  $-$15 04 28.60(0.09)  &   6.4(0.4)  &   7.1(0.6)  &  7.18  &  4.09  & 103.4 &    5.0 &     0.8 &     0.9 &     1.3  \\
J173554.6$-$150130              &  17 35 54.620(0.020) &  $-$15 01 30.50(0.10)  &   4.4(0.3)  &   4.3(0.5)  &  6.26  &  3.56  & 104.0 &    1.6 &     0.3 &     0.5 &     0.1  \\
J173559.4$-$143458              &  17 35 59.420(0.050) &  $-$14 34 58.80(0.50)  &   3.1(0.5)  &   5.0(2.0)  & 10.89  &  5.61  &  57.8 &   11.2 &     1.3 &     2.8 &     0.2  \\
J173607.7$-$144141              &  17 36 07.730(0.020) &  $-$14 41 41.00(0.30)  &   4.2(0.5)  &   2.9(0.6)  &  5.73  &  3.77  &  31.0 &    8.7 &     1.1 &     0.9 &     0.0  \\
J173616.6$-$150431              &  17 36 16.675(0.007) &  $-$15 04 31.36(0.06)  &  10.0(0.4)  &  10.1(0.6)  &  7.13  &  3.71  & 113.9 &    7.3 &     1.8 &     1.8 &     0.5  \\
J173623.7$-$152405              &  17 36 23.770(0.040) &  $-$15 24 05.00(0.60)  &   2.3(0.4)  &   4.0(2.0)  &  7.20  &  6.52  &  47.8 &    5.6 &     0.5 &     2.1 &     0.8  \\
J173632.5$-$144714              &  17 36 32.500(0.050) &  $-$14 47 14.70(0.60)  &   1.7(0.4)  &   2.5(0.8)  &  7.80  &  5.04  &  46.4 &    8.4 &     0.5 &     1.0 &     0.8  \\
J173638.8$-$152132              &  17 36 38.800(0.020) &  $-$15 21 32.70(0.20)  &   4.2(0.4)  &   4.6(0.7)  &  6.06  &  3.44  & 142.9 &    1.2 &     0.4 &     0.7 &     1.4  \\
J173640.5$-$144401              &  17 36 40.510(0.080) &  $-$14 44 01.00(1.00)  &   0.8(0.4)  &   1.6(0.8)  &  8.18  &  5.13  &  47.6 &    4.8 &     0.4 &     0.8 &     0.7  \\
J173645.6$-$151019              &  17 36 45.694(0.004) &  $-$15 10 19.09(0.06)  &  21.6(0.6)  &  19.5(0.8)  &  6.52  &  4.02  & 152.1 &   10.5 &     7.3 &     6.6 &     0.1  \\
J173651.0$-$144550              &  17 36 51.066(0.008) &  $-$14 45 50.10(0.20)  &   5.6(0.4)  &   5.6(0.6)  &  6.23  &  3.89  & 168.6 &    6.9 &     1.0 &     1.1 &     1.8  \\
J173651.2$-$145904              &  17 36 51.220(0.050) &  $-$14 59 04.30(0.40)  &   4.0(0.9)  &   5.0(2.0)  &  7.62  &  3.24  &  85.6 &    9.9 &     1.5 &     2.5 &     1.3  \\
J173700.7$-$145600              &  17 37 00.710(0.020) &  $-$14 56 00.70(0.20)  &  11.4(0.8)  &  13.0(2.0)  &  5.68  &  3.45  & 108.0 &    8.6 &     2.7 &     3.5 &     1.4  \\
J173703.3$-$144245              &  17 37 03.330(0.070) &  $-$14 42 45.00(0.70)  &   1.4(0.4)  &   3.0(1.0)  &  9.73  &  5.89  &  68.2 &    6.2 &     0.4 &     1.1 &     0.4  \\
J173704.3$-$153301$^{\rm (D)}$  &  17 37 04.361(0.001) &  $-$15 33 01.02(0.02)  & 141.2(0.5)  & 119.5(0.7)  &  7.51  &  3.22  & 150.2 &   13.3 &    89.7 &    75.9 &     0.4  \\
J173705.0$-$153300$^{\rm (D)}$  &  17 37 05.020(0.002) &  $-$15 33 00.78(0.03)  &  48.5(0.5)  &  40.5(0.7)  &  7.70  &  3.10  & 148.6 &   13.4 &    31.3 &    26.1 &     0.3  \\
J173714.2$-$150308              &  17 37 14.220(0.020) &  $-$15 03 08.30(0.20)  &   8.9(0.8)  &   9.0(2.0)  &  4.90  &  3.50  &  98.8 &    5.4 &     1.3 &     2.2 &     1.4  \\
J173716.3$-$144650              &  17 37 16.340(0.020) &  $-$14 46 50.60(0.40)  &   4.1(0.5)  &   5.0(0.8)  &  8.16  &  4.57  & 165.1 &   11.2 &     1.7 &     2.1 &     0.9  \\
J173735.2$-$152741              &  17 37 35.223(0.008) &  $-$15 27 41.70(0.20)  &  12.8(0.6)  &  11.3(0.8)  &  7.67  &  3.64  & 138.4 &   14.5 &    10.8 &     9.6 &     0.2  \\
J173811.6$-$150301              &  17 38 11.609(0.001) &  $-$15 03 01.41(0.01)  & 460.8(0.9)  & 405.0(2.0)  &  5.85  &  2.84  & 120.4 &   10.0 &   139.4 &   122.5 &     0.9  \\
J173818.5$-$151318              &  17 38 18.552(0.007) &  $-$15 13 18.70(0.20)  &  34.0(2.0)  &  46.0(3.0)  &  8.80  &  4.34  & 138.9 &   17.4 &    70.1 &    94.8 &     0.2  \\
J174358.1$-$100631              &  17 43 58.151(0.005) &  $-$10 06 31.27(0.06)  &  14.4(0.5)  &  16.2(0.8)  &  5.54  &  3.32  &  49.5 &    7.6 &     2.7 &     3.1 &     1.0  \\
J174400.5$-$101331              &  17 44 00.518(0.001) &  $-$10 13 31.99(0.02)  &  94.2(0.6)  &  81.6(0.8)  &  6.39  &  2.95  &  28.9 &   12.9 &    54.7 &    47.4 &     0.8  \\
J174441.9$-$101040$^{\rm (D)}$  &  17 44 41.974(0.001) &  $-$10 10 40.81(0.02)  & 106.3(0.7)  & 120.3(1.1)  &  6.16  &  4.42  & 178.1 &   10.1 &    32.9 &    37.3 &     0.6  \\
J174443.2$-$101001$^{\rm (D)}$  &  17 44 43.214(0.002) &  $-$10 10 01.71(0.05)  &  46.2(0.7)  &  77.6(1.4)  &  8.70  &  4.55  & 174.6 &    9.7 &    13.1 &    22.1 &     0.6  \\
J174457.8$-$101206$^{\rm (D)}$  &  17 44 57.865(0.002) &  $-$10 12 06.65(0.05)  &  33.8(0.4)  &  60.3(1.0)  &  8.79  &  5.34  & 175.5 &   11.0 &    12.6 &    22.5 &     1.4  \\
J174457.9$-$101228$^{\rm (D)}$  &  17 44 57.929(0.003) &  $-$10 12 28.39(0.07)  &  19.9(0.4)  &  30.7(0.9)  &  7.88  &  5.06  & 179.5 &   10.6 &     6.9 &    10.7 &     0.7  \\
J174500.3$-$093825$^{\rm (D)}$  &  17 45 00.356(0.004) &  $-$09 38 25.67(0.09)  &   6.8(0.2)  &  15.9(0.6)  &  7.59  &  4.60  &   1.0 &    3.7 &     0.5 &     1.3 &     1.5  \\
J174501.4$-$093849$^{\rm (D)}$  &  17 45 01.425(0.004) &  $-$09 38 49.39(0.06)  &   7.1(0.2)  &  15.0(0.5)  &  6.04  &  5.12  &  36.9 &    3.3 &     0.5 &     1.0 &     1.7  \\
J174507.2$-$093754              &  17 45 07.236(0.004) &  $-$09 37 54.11(0.06)  &   6.1(0.2)  &   6.2(0.4)  &  4.38  &  3.34  &  26.5 &    3.3 &     0.4 &     0.6 &     0.5  \\
J174519.2$-$092713              &  17 45 19.247(0.002) &  $-$09 27 13.67(0.05)  &  19.5(0.3)  &  18.9(0.5)  &  6.48  &  3.77  &  20.0 &   13.9 &    14.4 &    13.9 &     0.5  \\
J174520.3$-$095323              &  17 45 20.327(0.002) &  $-$09 53 23.23(0.04)  &  24.3(0.3)  &  20.9(0.4)  &  6.24  &  3.39  & 171.6 &   12.6 &    13.0 &    11.2 &     0.7  \\
J174528.1$-$102148              &  17 45 28.190(0.030) &  $-$10 21 48.40(0.40)  &   2.1(0.4)  &   4.3(0.9)  &  7.09  &  5.90  & 116.3 &    4.2 &     0.4 &     1.0 &     2.2  \\
J174535.5$-$101439              &  17 45 35.521(0.007) &  $-$10 14 39.00(0.20)  &  15.4(0.4)  &  56.6(1.6)  & 11.43  &  8.34  & 140.8 &    9.9 &     4.6 &    17.0 &     1.8  \\
J174539.0$-$093914              &  17 45 39.095(0.005) &  $-$09 39 14.99(0.07)  &   6.0(0.3)  &   5.9(0.4)  &  4.66  &  3.89  &  41.9 &    7.1 &     1.0 &     1.0 &     0.2  \\
J174548.3$-$102203              &  17 45 48.360(0.020) &  $-$10 22 03.80(0.40)  &   3.3(0.5)  &   3.8(0.8)  &  6.44  &  5.24  & 178.1 &    9.2 &     1.0 &     1.3 &     0.7  \\
J174552.6$-$100526              &  17 45 52.692(0.004) &  $-$10 05 26.63(0.06)  &   6.8(0.3)  &   6.9(0.5)  &  4.12  &  2.99  &  33.5 &    4.1 &     0.6 &     0.7 &     1.0  \\
J174556.0$-$100613$^{\rm (D)}$  &  17 45 56.043(0.006) &  $-$10 06 13.10(0.20)  &   4.1(0.3)  &   4.1(0.5)  &  3.99  &  3.15  &  24.7 &    4.5 &     0.5 &     0.6 &     0.6  \\
J174557.9$-$100608$^{\rm (D)}$  &  17 45 57.970(0.002) &  $-$10 06 08.93(0.03)  &  14.9(0.3)  &  15.5(0.5)  &  4.10  &  3.18  &  26.5 &    4.4 &     1.3 &     1.4 &     0.8  \\
J174604.5$-$093903              &  17 46 04.582(0.003) &  $-$09 39 03.95(0.03)  &  25.4(0.5)  &  23.0(0.6)  &  5.35  &  3.50  &  80.7 &   11.0 &     9.6 &     8.7 &     1.6  \\
J174605.7$-$095517              &  17 46 05.775(0.001) &  $-$09 55 17.51(0.02)  &  36.9(0.3)  &  45.5(0.6)  &  4.48  &  3.96  &  18.1 &    6.7 &     5.5 &     6.8 &     0.9  \\
J174613.9$-$093705              &  17 46 13.931(0.005) &  $-$09 37 05.15(0.06)  &  12.0(0.5)  &  10.3(0.6)  &  4.52  &  3.57  &  90.5 &    9.4 &     3.3 &     2.8 &     1.4  \\
J174616.5$-$102358$^{\rm (MD)}$ &  17 46 16.519(0.005) &  $-$10 23 58.12(0.06)  &   8.2(0.3)  &   8.6(0.4)  &  6.28  &  3.83  &  55.9 &    8.1 &     1.7 &     1.8 &     0.4  \\
J174624.1$-$095208$^{\rm (D)}$  &  17 46 24.147(0.003) &  $-$09 52 08.90(0.06)  &  16.2(0.4)  &  19.6(0.7)  &  5.83  &  3.96  &  25.4 &   11.3 &     6.6 &     7.9 &     0.5  \\
J174624.1$-$095201$^{\rm (D)}$  &  17 46 24.166(0.003) &  $-$09 52 01.06(0.06)  &  19.0(0.4)  &  23.0(0.7)  &  6.68  &  3.49  &  27.5 &   11.5 &     7.9 &     9.6 &     0.4  \\
J174625.4$-$101722              &  17 46 25.480(0.020) &  $-$10 17 22.30(0.20)  &   2.5(0.3)  &   1.9(0.4)  &  6.04  &  3.47  &  57.1 &    7.7 &     0.6 &     0.5 &     2.9  \\
J174630.2$-$100809              &  17 46 30.285(0.007) &  $-$10 08 09.10(0.20)  &   6.2(0.5)  &   4.4(0.5)  &  4.47  &  3.27  & 148.0 &    9.8 &     1.9 &     1.4 &     1.2  \\
J174631.8$-$103043              &  17 46 31.890(0.010) &  $-$10 30 43.40(0.20)  &   3.7(0.3)  &   3.3(0.4)  &  6.43  &  3.47  &  43.9 &    9.1 &     1.0 &     0.9 &     1.9  \\
J174639.2$-$102244              &  17 46 39.240(0.020) &  $-$10 22 44.00(0.20)  &   1.6(0.2)  &   1.7(0.4)  &  6.11  &  3.24  &  55.5 &    2.4 &     0.2 &     0.4 &     1.7  \\
J174726.2$-$102609              &  17 47 26.276(0.007) &  $-$10 26 09.15(0.07)  &   7.4(0.3)  &   6.4(0.4)  &  6.17  &  4.10  &  71.3 &    9.9 &     2.2 &     1.9 &     0.5  \\
J174727.8$-$095917$^{\rm (D)}$  &  17 47 27.888(0.001) &  $-$09 59 17.28(0.01)  & 129.4(0.4)  & 134.8(0.6)  &  3.51  &  2.81  &  36.5 &    3.4 &     8.4 &     8.7 &     1.6  \\
J174727.2$-$095911$^{\rm (D)}$  &  17 47 27.225(0.003) &  $-$09 59 11.51(0.04)  &  12.1(0.4)  &  13.3(0.6)  &  3.32  &  3.19  &  44.4 &    3.6 &     0.9 &     1.1 &     2.6  \\
J174731.3$-$095153              &  17 47 31.324(0.002) &  $-$09 51 53.58(0.04)  &  29.1(0.5)  &  27.9(0.7)  &  4.70  &  3.15  &   3.2 &   10.0 &     8.8 &     8.5 &     0.8  \\
J174731.3$-$101448              &  17 47 31.352(0.002) &  $-$10 14 48.06(0.02)  &  29.8(0.3)  &  29.9(0.4)  &  7.43  &  3.59  &  58.9 &   13.0 &    17.7 &    17.8 &     0.9  \\
J174735.0$-$100648              &  17 47 35.044(0.009) &  $-$10 06 48.00(0.20)  &   6.9(0.4)  &  26.9(1.5)  &  8.39  &  4.66  &  40.3 &    5.1 &     0.8 &     3.1 &     0.2  \\
J174742.8$-$095447              &  17 47 42.891(0.007) &  $-$09 54 47.80(0.20)  &   5.2(0.4)  &   4.5(0.6)  &  3.82  &  2.61  &  30.3 &    7.1 &     0.9 &     1.0 &     1.2  \\
J174805.4$-$100433              &  17 48 05.408(0.008) &  $-$10 04 33.30(0.20)  &   6.3(0.5)  &   5.5(0.6)  &  3.71  &  3.23  & 112.6 &    7.4 &     1.2 &     1.1 &     1.2  \\
J180809.1$-$104031$^{\rm (MD)}$ &  18 08 09.120(0.020) &  $-$10 40 31.00(0.50)  &   3.3(0.4)  &   8.0(2.0)  & 10.73  &  5.27  & 162.3 &    9.3 &     0.9 &     2.9 &     2.0  \\
J180809.6$-$101307              &  18 08 09.640(0.020) &  $-$10 13 07.50(0.20)  &  21.3(0.8)  &  63.0(3.0)  & 11.79  &  7.58  &  54.8 &   16.4 &    31.4 &    92.8 &     2.0  \\
J180819.6$-$105916              &  18 08 19.610(0.010) &  $-$10 59 16.80(0.10)  &  20.7(0.6)  &  24.0(1.0)  &  7.54  &  3.10  &  55.2 &   15.1 &    20.8 &    24.2 &     0.5  \\
J180821.5$-$105044              &  18 08 21.550(0.010) &  $-$10 50 44.10(0.20)  &  14.3(0.8)  &  19.0(2.0)  &  7.41  &  4.56  &  62.8 &   13.6 &     9.7 &    13.0 &     0.9  \\
J180824.7$-$104155              &  18 08 24.757(0.003) &  $-$10 41 55.30(0.10)  &  17.1(0.4)  &  16.7(0.6)  &  6.40  &  3.69  &   7.3 &    9.6 &     4.7 &     4.6 &     1.1  \\
J180834.3$-$103024$^{\rm (MD)}$ &  18 08 34.333(0.007) &  $-$10 30 24.90(0.40)  &   5.3(0.3)  &  19.0(2.0)  & 15.89  &  4.25  & 174.3 &    2.6 &     0.4 &     2.2 &     0.9  \\
J180902.8$-$105602              &  18 09 02.834(0.002) &  $-$10 56 02.71(0.02)  &  24.0(0.4)  &  26.2(0.6)  &  4.85  &  2.31  &  50.3 &    4.3 &     2.1 &     2.3 &     1.3  \\
J180903.0$-$101736              &  18 09 03.085(0.001) &  $-$10 17 36.23(0.02)  &  71.6(0.5)  &  72.1(0.7)  &  6.41  &  2.52  &  51.1 &    6.8 &    11.0 &    11.1 &     2.3  \\
J180908.6$-$101936              &  18 09 08.620(0.004) &  $-$10 19 36.63(0.05)  &  24.5(0.6)  &  23.8(0.8)  &  6.42  &  3.02  &  47.7 &    8.2 &     5.1 &     5.0 &     1.7  \\
J180910.2$-$103217              &  18 09 10.250(0.020) &  $-$10 32 17.90(0.30)  &   5.8(0.6)  &   4.9(0.7)  &  6.26  &  4.52  &  23.0 &   10.6 &     2.1 &     1.8 &     1.6  \\
J180916.6$-$110041              &  18 09 16.639(0.002) &  $-$11 00 41.54(0.04)  &  39.6(0.5)  &  71.0(2.0)  &  7.51  &  3.68  &  31.1 &    7.5 &     7.1 &    12.9 &     1.3  \\
J180942.8$-$100745              &  18 09 42.840(0.020) &  $-$10 07 45.00(0.20)  &   3.9(0.5)  &   4.0(0.7)  &  6.44  &  2.41  &  52.6 &    7.6 &     0.9 &     1.0 &     0.6  \\
J180943.4$-$104055$^{\rm (D)}$  &  18 09 43.400(0.020) &  $-$10 40 55.10(0.30)  &   3.6(0.4)  &   9.0(2.0)  &  7.09  &  4.97  &  37.2 &   10.0 &     1.2 &     3.4 &     $-$  \\ 
J180944.2$-$095755              &  18 09 44.295(0.008) &  $-$09 57 55.30(0.20)  &  21.9(0.7)  &  35.0(2.0)  & 12.31  &  3.40  &   3.4 &   15.4 &    23.6 &    37.8 &     0.1  \\
J180944.7$-$104045$^{\rm (D)}$  &  18 09 44.790(0.020) &  $-$10 40 45.40(0.30)  &   4.4(0.4)  &   8.8(0.9)  &  6.90  &  3.83  &  22.6 &    9.7 &     1.3 &     2.7 &     2.0  \\
J180945.1$-$100121              &  18 09 45.150(0.030) &  $-$10 01 21.80(0.30)  &   4.8(0.6)  &   8.0(2.0)  &  7.61  &  4.66  &  53.3 &   12.5 &     2.6 &     4.7 &     0.8  \\
J180948.8$-$103825              &  18 09 48.832(0.002) &  $-$10 38 25.20(0.03)  &  21.6(0.4)  &  23.5(0.6)  &  4.61  &  2.63  &  34.5 &    7.2 &     3.6 &     4.0 &     0.4  \\
J180954.1$-$104128              &  18 09 54.171(0.007) &  $-$10 41 28.10(0.20)  &   4.8(0.4)  &   4.2(0.6)  &  4.32  &  2.68  &  13.2 &    9.4 &     1.4 &     1.3 &     0.1  \\
J180955.9$-$110502              &  18 09 55.930(0.004) &  $-$11 05 02.30(0.10)  &  34.6(1.0)  &  30.0(2.0)  &  6.69  &  4.11  & 157.5 &   15.4 &    37.7 &    32.7 &     0.2  \\
J180957.9$-$103722              &  18 09 57.909(0.005) &  $-$10 37 22.80(0.10)  &   6.9(0.3)  &  16.4(0.9)  &  5.87  &  3.98  &   2.2 &    5.3 &     0.8 &     2.0 &     1.1  \\
J180959.3$-$110028              &  18 09 59.359(0.003) &  $-$11 00 28.31(0.05)  &  47.1(0.8)  &  68.0(2.0)  &  6.45  &  5.99  &  70.7 &   12.9 &    27.2 &    39.3 &     0.2  \\
J181004.6$-$102820              &  18 10 04.688(0.004) &  $-$10 28 20.29(0.06)  &   6.8(0.3)  &   8.8(0.6)  &  4.27  &  2.85  &  34.3 &    4.1 &     0.6 &     0.9 &     2.1  \\
J181011.2$-$104116              &  18 10 11.201(0.002) &  $-$10 41 16.31(0.03)  &  41.5(0.4)  &  70.9(0.9)  &  6.40  &  3.84  &   2.7 &    9.0 &    10.2 &    17.5 &     1.5  \\
J181011.9$-$101043              &  18 10 11.939(0.009) &  $-$10 10 43.90(0.08)  &   8.4(0.5)  &  10.3(0.7)  &  6.57  &  3.41  &  72.3 &   10.3 &     2.7 &     3.4 &     0.1  \\
J181017.4$-$102924$^{\rm (D)}$  &  18 10 17.405(0.005) &  $-$10 29 24.53(0.08)  &   5.4(0.3)  &   6.5(0.5)  &  4.36  &  2.56  &  36.6 &    4.3 &     0.5 &     0.7 &     0.6  \\
J181017.7$-$102907$^{\rm (D)}$  &  18 10 17.727(0.001) &  $-$10 29 07.80(0.01)  &  40.8(0.3)  &  41.0(0.5)  &  3.94  &  2.40  &  36.3 &    4.6 &     3.7 &     3.7 &     1.5  \\
J181030.9$-$101839$^{\rm (D)}$  &  18 10 30.980(0.030) &  $-$10 18 39.30(0.50)  &   2.5(0.3)  &  10.0(2.0)  &  9.70  &  5.43  &  30.8 &    9.0 &     0.7 &     3.2 &     2.6  \\
J181031.5$-$101822$^{\rm (D)}$  &  18 10 31.540(0.040) &  $-$10 18 22.40(0.60)  &   1.7(0.3)  &   7.0(2.0)  &  8.88  &  5.51  &  34.5 &    8.7 &     0.5 &     2.6 &     $-$  \\ 
J181040.2$-$103924              &  18 10 40.286(0.006) &  $-$10 39 24.27(0.09)  &  17.3(0.5)  &  18.5(0.8)  &  8.80  &  4.22  &  39.7 &   11.2 &     6.8 &     7.3 &     0.8  \\
J181058.2$-$100945              &  18 10 58.252(0.002) &  $-$10 09 45.42(0.04)  &   9.3(0.3)  &  10.7(0.4)  &  3.95  &  2.56  &  35.3 &    2.2 &     0.5 &     0.6 &     0.5  \\
J181115.0$-$110538              &  18 11 15.053(0.002) &  $-$11 05 38.19(0.04)  &  58.4(0.8)  &  45.8(0.9)  &  6.29  &  3.71  & 166.9 &   13.5 &    38.9 &    30.6 &     0.1  \\
J181123.4$-$101642              &  18 11 23.480(0.020) &  $-$10 16 42.60(0.20)  &   3.2(0.4)  &   2.6(0.5)  &  3.96  &  3.53  & 140.4 &    9.0 &     0.9 &     0.8 &     1.4  \\
J181125.7$-$104911              &  18 11 25.762(0.005) &  $-$10 49 11.08(0.06)  &  12.3(0.5)  &  12.4(0.7)  &  5.46  &  3.07  &  56.1 &    8.8 &     3.0 &     3.0 &     2.5  \\
J181136.3$-$101755              &  18 11 36.364(0.004) &  $-$10 17 55.77(0.06)  &  15.7(0.5)  &  11.9(0.6)  &  4.95  &  3.42  & 132.5 &   12.3 &     7.8 &     6.0 &     0.2  \\
J181138.5$-$102122              &  18 11 38.511(0.003) &  $-$10 21 22.44(0.06)  &  15.7(0.4)  &  13.2(0.5)  &  5.52  &  3.53  &   7.2 &   11.1 &     6.0 &     5.0 &     0.2  \\
J181149.9$-$102831              &  18 11 49.990(0.007) &  $-$10 28 31.50(0.20)  &   3.8(0.3)  &   3.9(0.5)  &  5.04  &  3.27  &  23.6 &    4.3 &     0.4 &     0.6 &     0.1  \\
J181209.7$-$104257              &  18 12 09.703(0.003) &  $-$10 42 57.76(0.06)  &  20.2(0.5)  &  16.1(0.6)  &  6.22  &  3.66  & 165.3 &   12.5 &    10.7 &     8.5 &     0.4  \\
J181248.2$-$104044              &  18 12 48.259(0.006) &  $-$10 40 44.66(0.08)  &  22.3(0.7)  &  23.9(0.9)  &  8.22  &  5.50  & 124.6 &   18.2 &    62.2 &    66.6 &     0.1  \\
J190309.9$-$114458              &  19 03 09.993(0.004) &  $-$11 44 58.41(0.04)  &  35.5(0.8)  &  39.6(1.2)  &  6.75  &  3.81  &  73.9 &   12.6 &    19.1 &    21.4 &     1.6  \\
J190310.6$-$112243              &  19 03 10.640(0.030) &  $-$11 22 43.90(0.40)  &   2.3(0.4)  &   2.3(0.7)  &  6.66  &  3.45  &  48.1 &    2.5 &     0.4 &     0.7 &     $-$  \\ 
J190318.8$-$110506              &  19 03 18.850(0.020) &  $-$11 05 06.40(0.40)  &   5.0(0.5)  &   9.6(1.2)  & 10.10  &  3.80  &  38.5 &   10.5 &     1.8 &     3.4 &     1.2  \\
J190325.6$-$113607              &  19 03 25.698(0.005) &  $-$11 36 07.81(0.07)  &  31.1(0.6)  &  50.6(1.3)  &  9.36  &  6.83  &  44.6 &   11.4 &    12.8 &    20.8 &     1.2  \\
J190338.8$-$114758$^{\rm (D)}$  &  19 03 38.840(0.020) &  $-$11 47 58.00(0.20)  &   5.1(0.5)  &   6.0(0.9)  &  5.37  &  3.12  &  67.9 &    5.9 &     0.8 &     1.2 &     2.6  \\
J190339.9$-$114756$^{\rm (D)}$  &  19 03 39.915(0.005) &  $-$11 47 56.54(0.05)  &  15.8(0.5)  &  16.9(0.9)  &  5.09  &  2.96  &  60.2 &    5.6 &     1.9 &     2.2 &     1.8  \\
J190340.7$-$111358              &  19 03 40.797(0.006) &  $-$11 13 58.46(0.09)  &  11.0(0.5)  &  10.7(0.7)  &  6.17  &  2.95  &  46.1 &   10.8 &     4.0 &     3.9 &     0.8  \\
J190341.1$-$114628              &  19 03 41.120(0.030) &  $-$11 46 28.60(0.20)  &   5.3(0.5)  &  10.3(1.2)  &  9.24  &  2.94  &  62.7 &    5.0 &     0.7 &     1.6 &     $-$  \\ 
J190341.1$-$112310$^{\rm (D)}$  &  19 03 41.191(0.002) &  $-$11 23 10.59(0.03)  &  40.1(0.4)  &  60.4(0.9)  &  7.81  &  5.02  &  52.9 &    7.3 &     6.9 &    10.4 &     $-$  \\ 
J190341.9$-$112249$^{\rm (D)}$  &  19 03 41.900(0.020) &  $-$11 22 49.80(0.20)  &   7.8(0.4)  &  15.2(1.1)  &  7.99  &  6.37  &  66.4 &    7.6 &     1.5 &     3.0 &     $-$  \\ 
J190341.8$-$110831              &  19 03 41.841(0.003) &  $-$11 08 31.47(0.04)  &  15.0(0.4)  &  14.2(0.6)  &  5.55  &  2.23  &  51.2 &    6.3 &     2.1 &     2.1 &     0.0  \\
J190343.4$-$110513              &  19 03 43.426(0.004) &  $-$11 05 13.19(0.04)  &  14.0(0.4)  &  14.2(0.6)  &  5.62  &  2.33  &  50.5 &    4.5 &     1.3 &     1.4 &     0.2  \\
J190359.0$-$110013              &  19 03 59.090(0.030) &  $-$11 00 13.70(0.50)  &   3.2(0.4)  &  13.3(1.6)  & 13.20  &  4.49  &  45.8 &    4.2 &     0.5 &     1.9 &     0.4  \\
J190404.2$-$113216              &  19 04 04.244(0.003) &  $-$11 32 16.04(0.03)  &  39.2(0.6)  &  43.0(0.8)  &  6.06  &  3.28  &  64.7 &   13.3 &    25.0 &    27.5 &     0.7  \\
J190405.0$-$113624              &  19 04 05.080(0.010) &  $-$11 36 24.90(0.20)  &  10.0(0.7)  &  11.1(1.1)  &  5.18  &  4.46  &  30.3 &    9.6 &     2.9 &     3.3 &     1.6  \\
J190430.2$-$115246$^{\rm (D)}$  &  19 04 30.278(0.002) &  $-$11 52 46.99(0.02)  &  90.1(0.8)  &  92.3(1.1)  &  5.17  &  4.45  &  98.7 &    9.8 &    26.3 &    26.9 &     0.7  \\
J190430.2$-$115241$^{\rm (D)}$  &  19 04 30.274(0.002) &  $-$11 52 41.07(0.02)  &  95.1(0.8)  &  86.8(1.0)  &  4.92  &  4.14  & 123.6 &    9.7 &    27.3 &    24.9 &     1.1  \\
J190431.8$-$112656              &  19 04 31.873(0.002) &  $-$11 26 56.69(0.02)  &  29.7(0.4)  &  28.7(0.5)  &  3.89  &  2.60  &  55.8 &    4.9 &     2.9 &     2.8 &     0.8  \\
J190440.2$-$110859              &  19 04 40.239(0.009) &  $-$11 08 59.20(0.10)  &  15.5(0.6)  &  32.4(1.5)  &  7.82  &  5.84  &  75.8 &   10.5 &     5.3 &    11.1 &     2.9  \\
J190501.2$-$112243              &  19 05 01.294(0.007) &  $-$11 22 43.16(0.09)  &   4.5(0.3)  &   4.6(0.5)  &  3.66  &  2.70  &  58.4 &    3.6 &     0.4 &     0.6 &     $-$  \\ 
J190506.7$-$110126              &  19 05 06.729(0.006) &  $-$11 01 26.56(0.07)  &  12.1(0.5)  &  11.6(0.7)  &  4.72  &  3.45  &  75.0 &    8.5 &     2.7 &     2.7 &     1.1  \\
J190512.6$-$112921              &  19 05 12.651(0.003) &  $-$11 29 21.37(0.04)  &  16.0(0.5)  &  13.7(0.6)  &  3.66  &  3.25  & 116.2 &    6.9 &     2.5 &     2.2 &     0.4  \\
J190523.0$-$110250$^{\rm (MD)}$ &  19 05 23.042(0.002) &  $-$11 02 50.46(0.02)  &  48.9(0.4)  & 101.3(0.9)  &  8.63  &  2.81  &  51.6 &    4.3 &     4.1 &     8.5 &     $-$  \\ 
J190528.6$-$115331              &  19 05 28.617(0.001) &  $-$11 53 31.89(0.01)  & 286.2(1.4)  & 264.2(2.0)  &  5.79  &  2.69  &  39.9 &    8.0 &    56.9 &    52.6 &     2.7  \\
J190530.8$-$112600              &  19 05 30.847(0.006) &  $-$11 26 00.90(0.06)  &  10.8(0.5)  &   8.5(0.6)  &  4.55  &  2.84  &  92.4 &    9.9 &     3.3 &     2.6 &     1.2  \\
J190531.0$-$112817              &  19 05 31.010(0.020) &  $-$11 28 17.90(0.20)  &   5.4(0.6)  &   5.3(0.8)  &  5.20  &  3.37  &  94.3 &   10.4 &     1.9 &     1.9 &     0.2  \\
J190544.6$-$105850              &  19 05 44.657(0.008) &  $-$10 58 50.10(0.20)  &   4.7(0.4)  &   4.8(0.6)  &  4.22  &  2.98  &  44.2 &    5.7 &     0.7 &     0.8 &     $-$  \\ 
J190555.5$-$105233              &  19 05 55.536(0.005) &  $-$10 52 33.25(0.08)  &  15.7(0.6)  &  16.2(0.8)  &  5.99  &  3.32  &  44.1 &   12.5 &     8.3 &     8.5 &     2.3  \\
J190601.7$-$112510$^{\rm (V)}$  &  19 06 01.781(0.009) &  $-$11 25 10.10(0.20)  &   7.7(0.6)  &   7.2(0.9)  &  5.00  &  2.77  &  53.6 &    6.5 &     1.3 &     1.4 &     3.0  \\
J190604.5$-$110408              &  19 06 04.571(0.004) &  $-$11 04 08.47(0.05)  &  11.6(0.4)  &  10.5(0.6)  &  4.08  &  2.86  &  59.3 &    6.2 &     1.6 &     1.5 &     1.8  \\
J190617.4$-$112850$^{\rm (V)}$  &  19 06 17.428(0.003) &  $-$11 28 50.11(0.04)  &  22.7(0.5)  &  24.6(0.8)  &  4.41  &  2.90  &  46.8 &    4.5 &     2.1 &     2.3 &     4.1  \\
J190640.7$-$112209              &  19 06 40.774(0.002) &  $-$11 22 09.86(0.05)  &  25.1(0.5)  &  47.3(1.1)  &  6.06  &  3.63  &   8.4 &    4.3 &     2.2 &     4.1 &     1.2  \\
J190644.8$-$111416$^{\rm (D)}$  &  19 06 44.842(0.003) &  $-$11 14 16.86(0.06)  &  33.6(0.7)  &  49.3(1.3)  &  6.54  &  4.31  &  19.0 &   11.6 &    14.5 &    21.3 &     1.3  \\
J190645.1$-$111434$^{\rm (D)}$  &  19 06 45.112(0.001) &  $-$11 14 34.16(0.02)  & 115.7(0.7)  & 167.6(1.3)  &  6.62  &  4.11  &  16.9 &   11.4 &    47.2 &    68.4 &     1.1  \\
\hline
\end{longtable}
{\normalsize
~\\
$^{\rm (D)}$ Component of an apparent double source (separation less than 45$^{\prime\prime}$).\\
$^{\rm (MD)}$ Marginally resolved double source. \\
$^{\rm (T)}$ Component of an apparent triple source.\\
$^{\rm (V)}$ Candidate variable radio source. \\
$^{\rm (*)}$ Extended component of a radio galaxy western lobe.\\
}
}


\begin{thebibliography}{}


\bibitem[Aharonian et al. 2005]{aharonian} Aharonian F. A., et al., 2005, Science, 309, 746

\bibitem[Albert et al. 2006]{albert} Albert, J., Aliu, E., Anderhub, H., et~al. 2006, Science, 312, 1771

\bibitem[Bondi et al. 2007]{bondi2007} Bondi, M., et al., 2007, A\&A, 463, 519

\bibitem[Bosch-Ramon et al. 2005]{br2005} Bosch-Ramon V., Romero, G. E., Paredes, J. M. 2005, A\&A, 429, 267

\bibitem[Bosch-Ramon et al. 2006a]{br2006a} Bosch-Ramon, V., Paredes, J. M., Romero, G. E., Torres, D. F. 2006a, A\&A, 446, 1081

\bibitem[Bosch-Ramon et al. 2006b]{br2006b} Bosch-Ramon, V.,  Romero, G. E., Paredes, J. M. 2006b, A\&A, 447, 263

\bibitem[Combi et al. 2003]{combi} Combi, J. A., Romero, G. E., Paredes, J. M., Torres, D. F., Rib\'o, M., 2003, ApJ, 588, 731

\bibitem[Condon et al. 1998]{condon} Condon, J. J. et al. 1998, AJ, 115, 1693

\bibitem[de Vries et al. 2004]{vries} de Vries, W. H., Becker, R. H., White, R. L., Helfand, D. J., 2004, AJ, 127, 2565

\bibitem[Dhawan et al. 2006]{d2006} Dhawan, V., Mioduszewski, A., Rupen, M., 2006, PoS, Proceedings
of the VI Microquasar Workshop, p. 52.1

\bibitem[Dubus 2006]{dub2006} Dubus, G., 2006, A\&A, 2006, 456, 801

\bibitem[Gregory \& Taylor 1986]{gt} Gregory, P. C., Taylor, A. R., 1986, AJ, 92, 371

\bibitem[Griffith et al. 1994]{gr1994}  Griffith,  M. R., Wright A. E., Burke B. F., Ekers R. D., 1994, ApJSS, 90, 179

\bibitem[Kaufman-Bernad\'o et al. 2002]{bernado} Kaufman-Bernad\'o M. M., Romero G.E., \& Mirabel I. F., 2002 A\&A 385, L10 

\bibitem[Kniffen et al. 1997]{k1997} Kniffen, D. A., Alberts, W. C. K., Bertsch, D. L., et al., 1997, ApJ, 486, 126
 
\bibitem[Hartman et al. 1999]{hartman} Hartman, R. C., Bertsch, D.~L., \& Bloom, S.~D.~et~al. 1999, \apjs, 123, 79

\bibitem[Lasker et al. 1990]{lasker} Lasker, B. M., et al., 1990, AJ, 99, 2019

\bibitem[Mattox et al. 2001]{mattox} Mattox, J. R., Hartman, R. C., Reimer, O., 2001, ApJS 135, 155

\bibitem[Nolan et al. 2003]{nolan03} Nolan P., Tompkins W., Grenier I. \& Michelson P. 2003, ApJ, 597, 615

\bibitem[Orellana et al. 2007]{orellana07} Orellana, M., Bordas, P., Bosch-Ramon, V. Romero, G. E. Paredes, J. M., 2007, A\&A, 476, 9
	
\bibitem[Paredes et al. 2000]{paredes00} Paredes, J. M., Mart\'{\i}, J., Rib\'{o}, M., \& Massi, M. 2000, Science, 288, 2340
   
\bibitem[Paredes et al. 2005]{paredes05} Paredes, J. M., et al., 2005, Ap\&SS, 297, 223

\bibitem[Punsly et al. 2000]{punsly} Punsly, B., Romero, G. E., Torres, D. F., Combi, J. A., 2000, A\&A, 364, 552

\bibitem[Romero et al. 1999]{romero} Romero, G. E. et al., 1999, A\&A, 348, 868

\bibitem[Romero et al. 2003]{romero03} Romero, G. E., Torres, D. F., Kaufman-Bernad\'o, M. M., Mirabel, I. F., 2003, A\&A, 410, L1
	
\bibitem[Romero et al. 2007]{romero07} Romero, G. E., Okazaki, A. T., Orellana, M., Owocki, S. P., 2007, A\&A 474, 15

\bibitem[Sowards-Emmerd et al. 2004]{soward}  Sowards-Emmerd, D., Romani, R. W., Michelson, P. F., Ulvestad, J. S., 2004, ApJ, 609, 564

\bibitem[Skrutskie et al. 2006]{st2006} Skrutskie, M.~F., et al., 2006, AJ, 131, 1163

\bibitem[Torres et al. 2001a]{torres01a} Torres D. F., Romero G. E., Combi J. A., et al. 2001a, A\&A, 370, 468

\bibitem[Torres et al. 2001b]{torres01b} Torres D. F., Pessah M. E., \& Romero G. E. 2001b, Astronomische Nachrichten, 322, 223 


\end{thebibliography}
\end{document}